\documentclass [12pt]{article}
\RequirePackage[english]{babel}
\RequirePackage[cp1251]{inputenc}
\usepackage{amsmath,amssymb}

\begin{document}
\sloppy
\begin{center}

{\bf \Large{F.A. Gareev,  I.E. Zhidkova}\\
\vspace*{0.5cm}
 New Mechanism of Low Energy  Nuclear Reactions Using Superlow Energy
External Fields}\\
\vspace*{0.2cm}
{\sl Joint Institute for Nuclear Research, Dubna, Russia\\
 e-mail:gareev@thsun1.jinr.ru}\\
\end{center}

\section{Introduction}
 The  review of possible stimulation mechanisms of LENR (low energy nuclear reaction) is presented in \cite{2}.
We have concluded that transmutation of nuclei at low energies and
excess heat are possible in the framework of the known fundamental
physical laws – the universal resonance synchronization principle
\cite{1} and based on it different enhancement mechanisms of
reaction rates are responsible for these processes \cite{2}. The
excitation and ionization  of atoms may play the role of a trigger
for LENR. Superlow energy of external fields may stimulate LENR
\cite{3}. Investigation of this phenomenon requires  the knowledge
of different branches of science: nuclear and atomic physics,
chemistry and electrochemistry, condensed matter and solid state
physics,...
 The puzzle of poor reproducibility of experimental data is the fact that LENR occurs in open systems and
it is extremely sensitive to parameters of external fields and
systems. The classical reproducibility principle should be
reconsidered for LENR experiments. Poor reproducibility and
unexplained results do not mean that the experiment is wrong.  Our
main conclusion is:
 LENR may be understood in terms of the known fundamental laws without any violation of the basic physics.
 The fundamental laws of physics should be  the same in micro- and macrosystems.
Let us start with the description of  the hydrogen atom structure in different models

\subsection{The Hydrogen  Atom}
We will  describe very shortly  the structure of a hydrogen atom
using  standard basic physics that is well established, both
theoretically and experemintally in micro- and  macrosystems.

\subsection{The Bohr Model}

At the end of the 19th century it was established that the
radiation from hydrogen was emitted at specific quantized
frequencies. Niels Bohr developed the model to explain this
radiation using four postulates:

1. An electron in an atom moves in a circular orbit about the nucleus under the influence of the
Coulomb attraction between the electron and the nucleus, obeying the laws of classical mechanics.

2. Instead of the infinity of orbits which would be possible in
classical mechanics, it is only possible for an electron to move
in an orbit for which its orbital angular momentum $L$ is
integral multiple of $\hbar$:
                                    $$ L=n\hbar,\;\;n=1,2,3,…\eqno(1) $$

3. Despite the fact that it is constantly accelerating, an
electron moving in a such an allowed orbit does not radiate
electromagnetic energy. Thus, its total energy $E$ remains
constant.

4. Electromagnetic radiation is emitted if an electron, initially
moving in an orbit of total energy $E_{i}$, discontinuously
changes its motion so that it moves in an orbit of total energy
$E_{f}$.The frequency $\nu$ of the emitted radiation is equal to
the quantity

$$\nu_{if}=\frac{E_{i}-E_{f}}{h},\eqno(2)$$

where $h$ is Planck's constant.
 The electron is held in a stable
circular orbit around a nucleus. The Coulomb force is equal to the
centripetal force, according to Newton's second law

$$\frac{e^{2}}{r^{2}}=\frac{mv^{2}}{r},\eqno(3)$$
where $r$ is is the radius of the electron orbit, and $v$ is the
electron speed. The force is central; hence from the quantization
condition (1)

$$L=\mid \vec{r}*\vec{p}\mid=mvr=n\hbar.\eqno(4)$$

After solving equations (3) and (4) we have

$$v=\frac{e^{2}}{n\hbar},\;r=\frac{n^{2}\hbar^{2}}{me^{2}}=n^{2}a_{0}.\eqno(5)$$
>From equation (3) the kinetic energy is equal to

$$E_{k}=\frac{1}{2}mv^{2}=\frac{e^{2}}{2r},\eqno(6)$$
and hence the total energy

$$E=E_{k}+V=\frac{e^{2}}{2r}-\frac{e^{2}}{r}=-\frac{e^{2}}{2r}.\eqno(7)$$

Having $r$ from equation (5) one can write the expression for the
energy levels for  hydrogen atoms

$$E=-\frac{me^{4}}{2\hbar^{2}n^{2}};\eqno(8)$$
the same results were obtained further by quantum mechanics.

Using the angular momentum quantization condition $L=pr=nh/2\pi$
and Louis de Broglie's relationship $p=h/\lambda$  between
momentum and wavelength one can get

$$2\pi r=n\lambda. \eqno(9)$$

{\sl $\otimes$ It means that the circular Bohr orbit is an
integral number of the de Broglie wavelengths. The Bohr model is
actually only accurate for a one-electron system. }

\subsection{ The Hydrogen Atom in Classical Mechanics}

Is it possible to understand some properties of a hydrogen atom
from classical mechanics ? The Hamiltonian for a hydrogen atom is

$$H=\frac{m_p \dot{\vec{r_p}}\;^2}{2} +
\frac{m_e \dot{\vec{r_e}}\;^2}{2} - \frac{e^2}{ \mid \vec{r}_p -
\vec{r}_e \mid }.\eqno(10)$$

All notation is standard. The definition of center of mass is

$$m_{p}\vec{r}_{p}+m_{e}\vec{r}_{e}=0, \eqno(11) $$

and the relative distance between electron and proton is

$$\vec{r}=\vec{r}_p- \vec{r}_e. \eqno(12)$$

Equations (10)-(12) lead to the results:

$$\vec{r}_{p}=\frac{m_{e}}{m_{p}+m_{e}}\vec{r},\;\vec{r}_{e}=
-\frac{m_{p}}{m_{p}+m_{e}}\vec{r}, \eqno(13)$$

$$H=\frac{\mu \dot{{\vec
r}}\;^2}{2}-\frac{e^{2}}{r},\eqno(14)$$

where
$$\mu=\frac{m_{p}m_{e}}{m_{p}+m_{e}}.\eqno(15)$$

The Hamiltonian (14) coincides with the Hamiltonian for the
fictitious material point with reduced mass $\mu$ moving in the
external field $-e^{2}/r$. If we known the trajectory of this
fictitious particle $\vec{r}=\vec{r}(t)$ then we can reconstruct
the trajectories of electron and proton using equations (13)

$$\vec{r}_{p}(t)=\frac{m_{e}}{m_{p}+m_{e}}\vec{r}(t),\;\;\;
\vec{r}_{e}(t)=-\frac{m_{p}}{m_{p}+m_{e}}\vec{r}(t).\eqno(16)$$

It is evident from (16) that the proton and electron  move in the
opposite directions synchronously. So the motions of proton,
electron and their relative motion occur with  equal frequency

$$\omega_{p}=\omega_{e}=\omega_{\mu},\eqno(17)$$
over the closed trajectories scaling by the ratio

$$ \frac{v_{e}}{v_{p}}=\frac{m_{p}}{m_{e}},\;\frac{v_{e}}{v_{\mu}}=\frac{m_{\mu}}{m_{e}},\;
\frac{v_{\mu}}{v_{p}}=\frac{m_{p}}{m_{\mu}}.\eqno(18)$$

I.A. Schelaev \cite{SCH04} proved that the frequency spectrum of
any motion on ellipse contains only one harmonic.

We can get from (16) that
$$\vec{P}_{p}=\vec{P},\;\vec{P}_{e}=-\vec{P}, \eqno(18)$$

where -- $\vec{P}_{i}=m_{i}\vec{\dot{r}}_{i}$. All three impulses
are equal to each other in absolute value, which means the
equality of
$$\lambda_{D}(p)=\lambda_{D}(e)=\lambda_{D}(\mu)=h/P.\eqno(19)$$

Conclusion:\\
{\sl $\otimes \;\;\;$ Therefore, the motions of proton and
electron and their relative motion occur with the same FREQUENCY,
IMPULSE (linear momentum) and the de Broglie WAVELENGTH. All
motions are synchronized and self-sustained. Therefore the whole
system -hydrogen atom is nondecomposable to the independent
motions of proton and electron despite the fact that the kinetic
energies ratio of electron to proton one is small:

$$ \frac{E_{k}(e)}{E_{k}(p)}=4.46*10^{-4}.$$
It means that the nuclear and corresponding atomic processes must
be considered as a unified  entirely determined whole process.

For example, V.F. Weisskopf \cite{6} come to conclusion that the
maximum height $H$ of mountains in terms of the Bohr radius $a$ is
equal to

$$\frac{H}{a}=2.6*10^{14},$$
and water waves lengths $\lambda$ on the surface of a lake in
terms of the Bohr radius is equal to}

$$\frac{\lambda}{a}\approx 2\pi*10^{7}.$$

Let us introduce the quantity $f=rv$ which is the invariant of
motion, according to the second Keplers law, then

$$ \mu v=\frac{\mu vr}{r}=\frac{\mu f}{r},\eqno(20)$$
and we can rewrite  equation (14) in the following way

$$ H=\frac{\mu f^{2}}{2r^{2}}-\frac{e^{2}}{r}. \eqno(21)$$

We can obtain the minimal value of (21) by taking its first
derivative over $r$ and setting it equal to zero. The minimal
value occurs at

$$ r_{0}=\frac{\mu f^{2}}{e^{2}}, \eqno(22)$$
and the result is

$$ H_{min}=E_{min}=-\frac{e^{4}}{2 \mu f^{2}}.\eqno(23)$$

The values of invariant of motion  $\mu f$ (in MeV*s) can be
calculate from (23)if we require the equality of  $E_{min}$ to the
energy of ground  state of a hydrogen atom

$$ \mu f= \mu
vr=6.582118*10^{-22}=\hbar,\;\eqno(24)$$

Conclusion:\\
{\sl $\otimes \;\;\;$ The Bohr quantization conditions were
introduced as hypothesis. We obtain these conditions from a
classical Hamiltonian requiring its minimality. It is necessary to
strongly stress that no assumption was formulated about
trajectories of proton and electron. We reproduced exactly the
Bohr result and modern quantum theory. The Plank constant $\hbar$
is the Erenfest adiabatic invariant for a hydrogen atom: $\mu vr =
\hbar$.}

Let us briefly review our steps:

$\bullet$ We used a well established interaction between proton
and electron.

$\bullet$ We used a fundamental fact that the total energy=kinetic
energy+potential energy.

$\bullet$ We used the second Kepler law.

$\bullet$ We used usual calculus to determine the minimum values
of $H$.

$\bullet$ We required the equality of  $E_{min}$ to the energy of
ground state of hydrogen atom.

 Classical Hamiltonian + classical interaction between proton and
electron + classical second Kepler law + standard variational
calculus -- these well establish steps in macrophysics reproduce
exactly  results of the Bohr model and modern quantum theory
(Schrodinger equation) -- results of microphysics. We have not
done anything spectacular or appealed to any revolutionary and
breakthrough physics.

M. Gryzinski \cite{GRY04} has proved using the Newton equation
with well established interactions that atoms have the
quasi-crystal structure with definite angles: $90^{\circ}$,
$109^{\circ}$ and $120^{\circ}$, which are the well-known angles
in crystallography.

\section{Nuclei and Atoms as Open Systems}

1) LENR may be understood in terms of the known fundamental laws
without any violation of the basic physics. The fundamental laws
of physics should be  the same in micro- and macrosystems.

2)Weak and electromagnetic interactions may show a strong
influence of the surrounding conditions on the nuclear
processes.\\

3)The conservation laws are valid for  closed systems. Therefore,
the failure of parity in weak interactions means that the
corresponding systems are  open systems. Periodic variations (24
hours, 27, and 365 days in  beta-decay rates indicate that the
failure of parity in weak interactions has a cosmophysical origin.
Modern quantum theory is the theory for closed systems. Therefore,
it should be reformulated for open systems. The closed systems are
idealization of nature,  they do not exist in reality. \\

4)The universal resonance synchronization principle  is a key
issue to make a bridge between various scales of interactions and
it is responsible for self-organization of hierarchical  systems
independent of substance, fields,  and interactions. We give some
arguments in favor of the mechanism – ORDER BASED on ORDER,
declared  by Schrodinger in \cite{4}, a fundamental problem of
contemporary science.\\

5)The universal resonance synchronization principle became a
fruitful interdisciplinary science of general laws of
self-organized processes in different branches of physics because
it is the consequence of the energy conservation law and resonance
character of any interaction between wave systems. We have proved
the homology of  atom, molecule and crystal structures including
living cells. Distances of these systems are commensurable  with
the de Broglie wave length  of an electron in the ground state of
a hydrogen atom,  it plays the role of the standard distance, for
comparison. \\

6)First of all, the structure of a hydrogen atom should be
established. Proton and electron in a hydrogen atom move with the
same frequency that creates attractive forces between them, their
motions are synchronized. A hydrogen atom represents radiating and
accepting  antennas (dipole) interchanging  energies with the
surrounding  substance. The sum of radiate and absorb energy flows
by electron and  proton in a stable orbit is equal to zero
\cite{5} – the secret of success of  the Bohr model (nonradiation
of  the electron on stable orbit). “The greatness of mountains,
the finger sized drop, the shiver of a lake, and the smallness of
an atom are all related by simple laws of nature” – Victor F. Weisskopf  \cite{6}\\

7)These flows created  standing waves due to the resonance
synchronization principle. A constant energy exchange with
substances (with universes) create stable auto-oscillation systems
in which the frequencies of  external fields and all subsystems
are commensurable. The relict radiation (the relict isotropic
standing waves at T=2.725 K – the Cosmic Microwave Background
Radiation (CMBR))  and   many isotropic standing waves in cosmic
medium \cite{7} should be results of  self-organization of the
stable hydrogen atoms, according to the universal resonance
synchronization principle that is a consequence of  the
fundamental energy conservation law. One of the fundamental
predictions of the Hot Big Bang theory for the creation of the
Universe is CMBR.\\

8)The cosmic isotropic standing waves (many of them are not
discovered jet) should play the role of a conductor responsible
for stability of elementary particles, nuclei, atoms,…, galaxies
ranging in size more than 55 orders of magnitude.\\

9)The phase velocity of standing microwaves can be extremely high;
therefore all objects of the Universe should  get information from
each other almost immediately using phase velocity.

The aim of this paper is to discuss the possibility of inducing
and controlling nuclear reactions at low temperatures and
pressures by using different low energy external fields and
various physical and chemical processes. The main question is the
following: is it possible to enhance LENR rates by using  low and
extremely low energy external fields? The review of possible
stimulation mechanisms is presented in \cite{2,5}. We will discuss
new  possibilities to enhance LENR rates in condensed matter.

\section{LENR in Condensed Matter}

The modern understanding of the decay of the neutron is

$$n \rightarrow p+e^{-}+\overline{\nu_{e}}.\eqno(25)$$
The analysis of the energetics  of the decay can be performed
using the concept of  binding energy and the masses of  particles
by their rest mass energies. The energy balance from neutron decay
can be calculated from the particle masses. The rest mass
difference ( $0.7823 MeV/c^{2}$) between neutron and
(proton+electron) is converted to the kinetic energy of proton,
electron and neutrino. The neutron is about $0.2\%$ more massive
than a proton, an energy difference is 1.29 $MeV/c^{2}$. A free
neutron will decay with a half-life of about 10.3 minutes. Neutron
in a nucleus will decay if a more stable nucleus results otherwise
neutron in a nucleus will be stable. A half-life of neutron in
nuclei changes dramatically and depends on the isotopes.

It is possible the capture  electrons by protons

$$p+e^{-}\rightarrow n+\nu_{e},\eqno(26)$$
but for free protons and electrons this reaction has never been
observed which is the case in nuclear+ atomic physics. The capture
electrons by protons in a nucleus will occurs if a more stable
nucleus results. \\

\subsection{Cooperative Processes}

The processes (25) and (26) in LENR  are going with individual
nucleons and electrons. In these cases the rest mass difference is
equal to $0.7823 MeV/c^{2}$. In the case neutron decay
corresponding energy ($Q=0.7823$ MeV) converted to kinetic
energies of proton, electron and antineutrino. In the case capture
electrons by protons the quantity  $Q=0.7823$ MeV is a threshold
electron kinetic energy less which process (26) forbidden for free
proton and electron.

We have formulated the following postulate:\\
$\otimes$ {\sl The processes (25) and (26) in LENR  are going  in
the whole system: cooperative processes including all nucleons in
nuclei and electrons in atoms, in condensed matter. In these cases
a threshold energy $Q$ can be drastically decreased by internal
energy of whole system or even more -- the electron capture by
proton can be accompanied by emission of internal binding energy -
main source of excess heat phenomenon in LENR. }

The processes (25) and (26) are weak processes. A weak interaction
which is responsible for electron capture and other forms of beta
decay  is of a very short range. So the rate of electron capture
and emission (internal conversion) is proportional to the density
of electrons at the nuclei. It means that we can manage the
electron-capture (emission) rate by the change of the total
electron density at the
nuclei using different low energy external fields. These fields \\
can play a role of triggers for extracting internal energy of
whole system or subsystems, changing quantum numbers of the
initial states such way that forbidden transitions become allowed
ones. The distances between proton and electron in atoms are order
$10^{-6}--10^{-5}$ cm and any external field decreasing these
distances even for small value, can increase process (26) in
nuclei exponential way. Therefore, the influence of external
electron flux (discharge in condensed matter: breakdown, spark and
ark) on the velocities processes (25) and (26) can be great
importance.

The role of external electrons is the same as the catalytic role
of neutrons in the case of the chain fission reactions in nuclei
-- neutrons bring to nuclei binding energies (about 8 MeV) which
enhance the fission rates about 30 orders.

\section{Predicted Effects and Experimentum  Cruices}

Postulated enhancement mechanism of LENR by external fields can be
verified by the Exprimentum Cruices. We \cite{5} predicted that
natural geo-transmutation in the atmosphere and earth occur at the
the regions of strong change in geo-, bio-, acoustic-,...  and
electromagnetic fields.

The various electrodynamic processes at thunderstorms are
responsible for different phenomena: electromagnetic pulses,
$\gamma$-rays, electron fluxes, neutron fluxes, radioactive nuclei
fluxes.

\subsection{Production of Radiocarbon and Failing of Radiocarbon
Dating}

The half-life of radiocarbon $^{14}C$ is  5730 years. It was
developed a method for historical chronometry assuming that the
decay ratio of $^{14}C$ is constant on time. Radiocarbon dating is
widely used in archeology, geology, antiquities,...
 Let us consider the reaction

$$^{14}_{7}N+e^{-}\rightarrow ^{14}_{7}C+\nu_{e},\eqno(27)$$

the $T_{k}(e)$=156.41 keV is the threshold energy which should by
compared with 782.3 keV for process (26). Production of
radiocarbon by lighting bolts was established in \cite{LIB73}.
Unfortunately, this means failing of radiocarbon dating.

\subsection{Production Radiophosphorus by Thunderstorms}

The life-times of $^{32}_{15}P$ and $^{33}{15}P$ are equal to
14.36 and 25.34 days respectively. They were found in rain-water
after thunderstorms \cite{SEL70}. Production of the
radiophosphorus by thunderstorms can be understood by the
following way:

$$^{32}_{16}S+e^{-}\rightarrow ^{32}_{15}P+\nu_{e},\eqno{28}$$
$$^{33}_{16}S+e^{-}\rightarrow ^{33}_{15}P+\nu_{e},\eqno{29}$$
thresholds of these processes are equal to 1.710 and 0.240 MeV
respectively.  It is possible the precipitation of MeV electrons
from the inner radiation belt \cite{INAN} and enhancement the
processes by lighting.

\subsection{Neutron Production by Thunderstorms}

 Authors of paper \cite{BRA4} concluded that  neutron burst are
associated with lighting. The total number of neutrons produced by
one typical lighting discharge was estimated as $2.5*10^{10}$.

\subsection{LENR Stimulated by Condensed Matter Discharge}

Let us consider the discharge (breakdown, spark and arc) using the
different electrode.  There are the following processes:\\

1. The electrode is $Ni$. Orbital or external electron capture
$$^{58}_{28}Ni(68.27\%)+e^{-}\rightarrow ^{58}_{27}Co(70.78\;
days)+\nu_{e}, \eqno(30)$$

The threshold $Q_{1}=0.37766\;keV$ of this reaction on $Ni$ should
be compared with the threshold $Q=0.7823$ energy for electron
capture by free protons: $Q_{1}/Q_{2}\approx 2$. Velocity of
orbital electron capture can be enhanced by the discharge.

2.Orbital or external electron capture
$$^{58}_{27}Co(70.78\; days)+e^{-}\rightarrow ^{58}_{26}Fe(0.28\%)+\nu_{e},\eqno(31)$$
with emission of energy $Q_{2}=2.30408$ MeV.

3. Double orbital or external electron capture
$$^{58}_{28}Ni(68.27\%)+2e^{-} \rightarrow ^{58}_{26}Fe(0.28\%)+2\nu_{e},\eqno(32)$$
with emission of energy $Q_{3}=1.92642$

The proposed cooperative mechanism of LENR in this case can be
proved extremely simple way: presence of radioactive
$^{58}_{27}Co$ and enriched isotope of $^{58}_{26}Fe$.

$\otimes$ {\sl This mechanism can give the possibilities to get
controlling way the necessary isotopes and excess heat for using.}

\section{Conclusion}

We proposed a new mechanism of LENR: cooperative processes in
whole system - nuclei+atoms+condensed matter can occur at smaller
threshold then corresponding ones on free constituents. The
cooperative processes can be induced and enhanced by low energy
external fields. The excess heat is the emission of internal
energy and transmutations at LENR are the result of redistribution
inner energy of whole system.

\end{document}